------------------cut off here ---------------------

\documentstyle[12pt]{article}
\begin{document}

\begin{center}

{\Large {\bf A proposed gravitodynamic theory}}

\vskip 2cm
              T. Chang

\vskip 0.8cm

            Center for Space Plasma and Aeronomic Research\\
            University of Alabama in Huntsville\\
            Huntsville, AL 35899, USA\\

\vskip 2.0cm

{\bf Abstract}

\end{center}

\medskip
 This paper proposes a gravitodynamic theory  
 because there are similarities between 
gravitational theory and electrodynamics.
Based on Einstein's principle of equivalence,  
 two coordinate conditions are proposed into the four-dimensional
 line element and transformations.  As a consequence,
the equation of motion for  
gravitational force or inertial force has a
 form similar to the equation of Lorentz force on a charge 
in electrodynamics.  The inertial forces in  a
 uniformly rotating system 
are calculated, which show that the Coriolis force is
produced by a magnetic-type gravitational field. We have
also calculated the Sagnac effect due to the rotation.
These experimental facts strongly support our proposed coordinate
conditions. In addition,
the gravitodynamic field equations are briefly discussed.
 Since only four gravitational
potentials (3 + 1 split) enter the metric tensor, 
the gravitodynamic field equations in ``3+1 split" form 
would be analogous to Maxwell's equations.

\vskip 2.5cm

\begin{center}

January  1995 

\end{center}

\newpage

\noindent
{\bf 1. Introduction}
\vskip 0.1cm

In a recent review paper, Norton$^{[1]}$ has shown 
that the principle of general 
covariance has been disputed as physically vacuous. The debate over
this principle persists today.
In general relativity, space-time coordinates can be arbitrarily
chosen,$^{[2,3]}$ and gravitational potentials are represented 
by the metric tensor, $g_{{\mu}{\nu}}$, in the four-dimensional (4-D)
 line element: 

 $$ {ds}^2 = g_{{\mu}{\nu}}{ dx}{^{\mu}}{ dx}
{^{\nu}}     \eqno (1) $$

\noindent
where $g_{{\mu}{\nu}}$ has ten independent components,
and Greek letters $\mu$ and $\nu$ run from 0 to 3. In $Eq.$(1),
 {\it ds}$^2$ is an invariant under space-time coordinate
transformation.
 However, in order to get unique solutions of 
 the Einstein field equation,
certain coordinate conditions must be imposed. Ref.[4] is
a recent example that discusses the choice of coordinate conditions.
In this paper, we emphasize that constraints on  
coordinate conditions are imposed by the principle of equivalence.

Einstein illustrated the principle of equivalence using the following
simple example. Let $\Sigma$ be an inertial reference system; let another
system $K$ be uniformly accelerated with respect to $\Sigma$. 
Then relative to $K$, 
all free bodies have equal and parallel accelerations.
They behave just as if a gravitational field were present and
K were unaccelerated.$^{[2,3]}$ In another words, inertial force
is a type of experimentally confirmed ``gravitational force".

In order to formulate the above simple equivalent gravitational
field suggested by 
Einstein, let us start with an inertial system $\Sigma$. Let 
{\it X}$^\mu$ = ({\it cT,X,Y,Z}) = ({\it cT}, ${\bf R}$)
 be the pseudo-Cartesian coordinate in the
system $\Sigma$. Then the 4-D line element takes the form

\vskip 0.2cm
\hspace {1.5in}  {\it ds}$^2$ = ({\it cdT})$^2$ - {\it dX}$^2$ - 
{\it dY}$^2$  - {\it dZ}$^2$   \hspace {1.5in} (2)        

\vskip 0.2cm
 We now consider a system $K$ 
with coordinates {\it x}$^\mu$ = ({\it ct,x,y,z}) = 
({\it x$^o$}, {\bf r}),
which has a constant acceleration, ${\bf a}$, 
with respect to  $\Sigma$. Since the
Lorentz transformations are not valid between $\Sigma$  
and $K$, the following quasi-Galilean
transformations are usually introduced:$^{[5]}$
 $$  {\bf R} = {\bf r} + \frac{1}{2}{\bf a}t^2 ; 
\hskip 0.8cm T = t      \eqno (3) $$

\noindent
A simple calculation for Eqs.(2) and (3) gives
 $$  { ds}^2 = (1-({ at})^2/{ c}^2)({ dx^o})^2
  -2({\bf {\bf a}}t/{c})\cdot {d}{\bf r}{dx^o}
     - {\bf dr}^2     \eqno (4)  $$

\noindent
Comparing (4) with (1), we obtain $g_{oo}=(1-{a^2}{t^2}/{c^2})$; 
$g_{oi}=-{\it a_it/c}$; $g_{ij} = -{\delta_{ij}}$;
 Latin letters {\it i} and {\it j} run from 1 to 3.

We substitute these components of $g_{{\mu}{\nu}}$ into the
geodesic equation giving
 $$ \frac{ d^2  x^\mu}{d \tau^2} 
+ {\Gamma}{^\mu}_{\rho \lambda}
{\frac{d x^\rho}{d \tau}} {\frac{dx^\lambda}{d \tau}} = 0
  \eqno (5) $$

\noindent
In the case of low velocity approximation, $Eq.$(5) reduces to
${{\it d^2 \bf r}/{\it d t^2} = -{\bf a}}$.
This indicates that a particle with a rest mass {\it m}$_o$
in a uniformly accelerated system
should experience a uniformly inertial force ($-{\it m}_o${\bf a}).
 Although this result is well known,
the above calculation shows that certain constraints on 
space-time coordinates and transformations should be imposed by 
the principle of equivalence. As is well known,
the inertial force is a type of measurable force. Just like
other forces, inertial force is a vector, which can be reduced
or balanced by other forces.  However, if arbitrary curved
 coordinates are chosen, in general, the force components cannot be
singled out from the expression of $Eq$.(5). Then the physical meaning
would be lost in the mathematical maze. 
This is a major reason whythe debate on the principle 
of general covariance still persists.$^{[1]}$ 
Only when the equation of motion in a vector form is derived,
curvilinear spatial coordinates can be introduced for different
applications.

\vskip 0.6cm
\noindent
{\bf 2. Two Proposed Coordinate Conditions }

Based on the principle of equivalence,
 we propose two coordinate conditions into the 4-D
line element and transformations in our formulation: 

(i) In any reference system, the 4-D line element takes
the standard form
 $$  ds^2 = \Psi_o(dx^o)^2 - 2({\bf \Psi}\cdot 
d{\bf r})dx^o - d{\bf r}^2       \eqno (6) $$ 

\noindent 
Comparing $Eq.$(6) with {\it ds}$^2$ = 
{\it g}$_{\mu \nu}${\it dx}$^{\mu}${\it dx}$^{\nu}$, we have 
 $g_{oo} = \Psi_o $, $g_{oi} = -{\Psi_i} $,
$g_{ij} = -{\delta_{ij}}$. 

(ii) Between two reference systems, $K$ and $K'$, the value of 
{\it {ds}$^2$} is an invariant, but space-time coordinates
take infinitesimal Galilean transformations
 $$ {\bf dr}' = {\bf dr} - {\bf v}dt ;             
 \hspace {0.8 in}   t' = t   \eqno  (7) $$             

\noindent 
where ${\bf v}= {\bf v}({\bf r}$,{\it t}) is the 3-D 
relative velocity between two systems $K$ and $K'$. As a
special case, one reference 
system can be an inertial frame which has Minkowski metrics.

The physical implication of the above proposed conditions is 
described by Weinberg.$^{[6]}$  He 
has emphasized that a local gravitational field has two kinds of
sources.  One is nearby mass; another is all the mass
in the universe. It can be anticipated
that the inertial systems are determined by the mean background
gravitational field produced by all the matter of the universe.
This background gravitational field (or simply called ether)
provides a base for us to
propose the above two specific coordinate conditions.

Landau et al$^{[7]}$ and M\o ller$^{[5]}$ have rewritten 
the 4-D line element, $Eq.$(1), in another form:

  $$ { ds}^2 =  (\gamma_o{ dx}^o - \gamma_i { dx}^i)^2
 - {d}\sigma^2      \eqno(8)       $$ 

\noindent 
Comparing $Eq.$(8) with $Eq.$(6), 
 $\gamma_o$ = $\sqrt{{\it g}{_{oo}}}$ = $\sqrt{\Psi_o}$;
  $\gamma_i = -g_{oi}$/$\sqrt{{\it g}{_{oo}}}$ = 
$\Psi_i$/$\sqrt{\Psi_o}$; {\it d}$\sigma^2$ = ${\gamma _{ik}}$
{\it dx}${^i}${\it dx}${^k}$, where 
${\gamma _{ik}}=-g_{ik}$ + ${\gamma _i}{\gamma _{k}}$ = 
$\delta_{ik}$ + ${\Psi _i}{\Psi _k}/{\Psi_o}$.
 {\it d}$\sigma^2$ is the 3-D measured spatial interval in 
system $K$.  The proper time is defined as  
{\it d}{$\tau $ }= {\it ds}/c, which is an invariant and
indicates the measured time interval in system $K$. 
For slow motion and weak field approximation, 
 it is easily seen that the coordinate time interval
{\it dt}{$\simeq$}{$d\tau $ }, and the coordinate spatial interval
$d{\bf r}^2 \simeq${\it d}$\sigma^2$.

Notice that  $\Psi_o$ and $\bf {\Psi}$ are components of
 metric tensor g$_{\mu \nu}$ in $Eq.$(6). 
Although they have properties of gravitational
potentials, they do not form
a 4-D vector. From $Eq.$(6), it is easy to derive the transformations
of these potentials
 $$ {\Psi_o}' = {\Psi_o} - 
 \frac{v^2 }{{ c}^2} - {\bf \Psi}\cdot \frac{\bf v}{c} ;
\hspace {1cm} {\bf \Psi}' = {\bf \Psi} + \frac{\bf v}{c}  
   \eqno (9) $$

\noindent 
These transformations form a group with the parameters {\bf v}.

\vskip 0.6cm
\noindent 
{\bf 3. Gravitational Equation of Motion }
 
In this section, we will rewrite the general equation of motion, 
$Eq.$(5), into an explicit form in terms of the proposed two conditions.
We consider the variational principle $^{[7]}$
  $$  {\delta} {\int} (-m{_o}{ c}){ ds} = 
   {\delta} {\int} {L}({x}{^i},{u}{^i},{t}){dt} 
   \eqno  (10)    $$

\noindent
where {\it L}({\it x}{$^i$},{\it u}{$^i$},{\it t}) is the Lagrangian. 
Using $Eq.$ (6), we obtain
 $$ { L( x^i},{u^i},{t}) = (-m_o{c}^2){\sqrt{{\Psi_o} -  
2{\bf \Psi}\cdot {\bf u}/c  - u{^2}/c{^2}}} \eqno  (11) $$

\noindent
where {\bf u} = {\it d}{\bf r}/{\it dt} is the velocity of a particle
in system K.

Let ${\Psi_o}$ = (1 + 2$\phi$).  
Substituting the Lagrangian (11) into the Lagrange equation,$^{[7]}$

   $$  \frac{d}{dt}({\frac{\partial L}
{{\partial}{\bf u}}}) = { \frac{\partial L}{{\partial}{\bf r}}}
     \eqno (12)   $$

\vskip 0.2cm
\noindent 
we obtain the equation of motion:
     $$ {\frac{d(m {\bf u})}{dt}} + 
 c{\bf \Psi}{\frac {dm}{dt}} =  
m({\bf g} + \frac{\bf u}{c} \times {\bf h})    \eqno (13)  $$

\noindent 
where {\bf g} and {\bf h} are defined as
   $$ {\bf g} = c^2(-{\bigtriangledown}{\phi} -
{\frac{1}{c}}{\frac{\partial {\bf \Psi}}{\partial t}}) ; 
\hspace {1cm} {\bf h} = c^2(\bigtriangledown \times {\bf \Psi}) 
         \eqno (14)  $$

\noindent 
Here, {\bf g} is gravitational intensity and 
{\bf h} is magnetic-type gravitational intensity. 
Both of them have the units of {\it cm/sec}$^2$.
Therefore, the gravitational force is analogous to the
Lorentz force on a charged particle in an electromagnetic field.
As is discussed in section 1, force is a vector, which can 
be reduced or balanced by other forces. We now 
have clearly singled the force term out in the right hand side of
$Eq.$(13).  However, there are two different 
terms between the Lorentz formula and $Eq.$(13):
 (i) a moving mass {\it m} = {\it m}$_o \Gamma $, where
$\Gamma $ =  $({\Psi_o} -  
 2{\bf \Psi}\cdot {\bf u}/c  - u{^2}/c{^2})^{-1/2}$,
while the electric charge is a constant; (ii)
 a small additional term on the left hand side of $Eq.$(13),
 which is related to the changing rate of a moving mass.
We stress that $Eq.$(13) is the explicit formulation of the gravitational 
equation of motion, which is also valid to describe  
the magnetic-type gravitational force
as well as inertial force in non-inertial systems.

We now consider the total energy of a moving particle with a velocity 
{\bf u} in a gravitational field. According to Lagrange's 
formulation, the total energy of a particle is defined as$^{[7]}$
 
 $$     E_t = \frac{\partial L}{\partial {\bf u}} \cdot {\bf u} 
 - L     \eqno (15)         $$

\noindent
Substituting the Lagrangian expression (11) into (15), one obtains

  $$   E_t = m_o{\Gamma}c^2 [(1 + 2\phi) - 
{\bf \Psi}\cdot{\bf u}/c]    \eqno (16)       $$

\noindent
For a weak field, $\phi, \Psi \ll$ 1, an approximation of $Eq.$(16)
is
 
  $$   E_t = \frac{m_o c^2 (1 + \phi)}{\sqrt{1-u^2/c^2}} 
 \eqno (17)       $$

\noindent
$Eq.$(17) is essentially the same as the formula derived by Landau and
Lifshitz.$^{[7]}$

\vskip 0.6cm
\noindent
{\bf 4. Inertial Force in Accelerated Systems} 

 In terms of the formulation described above, inertial force 
is a type of experimentally confirmed gravitational force. 
We now calculate two typical cases of inertial force.

(a). A Uniformly Linear Accelerated System

 In the introduction of this paper,
we discussed the inertial force in
a uniformly linear accelerated system.
From $Eqs$.(4) and (6), we have $\phi$ = 
$-{a^2t^2}/2{c^2}$, ${\bf \Psi}$ = ${\bf a}t/c$. 
By using $Eq.$(14), we obtain
   $$ {\bf g} =  - c{\frac{\partial {\bf \Psi}}
{\partial t}} = - {\bf a} ;  \hskip 0.8cm
{\bf h} = 0   \eqno (18)  $$

\noindent
Notice that this equivalent gravitational intensity {\bf g} in 
$Eq.$(18) is produced by the 3-D gravitational vector potential.

(b). A Uniformly Rotating System

Suppose a uniformly rotating system has a constant angular
velocity $\bf \Omega$ = $\Omega {\bf k}$ with respect to an inertial
system $\Sigma$. Let the infinitesimal 
Galilean transformation be

 \hspace {0.7 in} {\bf dr} = {\bf dR} $-({\bf \Omega}\times$ {\bf R})
{\it dT} ; \hspace {0.7 in} {\it t} = {\it T} 
               \hspace {1.6 in}  (19)              

\noindent
Substituting $Eq.$(19) into $Eq.$(2), we obtain
the 4-D line element in the rotating system $K$ as follows:

 $$  {ds}^2 = (1 - ({\bf \Omega} \times 
{\bf r})^2/{c}^2)({dx^o})^2
  - \frac{2}{c}({\bf \Omega}\times {\bf r}) \cdot {d}{\bf r}
{dx^o} - {\bf dr}^2     \eqno (20)  $$

\noindent
From $Eq.$(20), we have $\phi$ = $-({\bf \Omega}\times 
{\bf r})^2$/(2{\it c}$^2$), ${\bf \Psi}$ = 
  $({\bf \Omega}\times {\bf r})/{\it c}$. Using $Eq.$(14), we obtain
     
 $$  {\bf g} = \frac{1}{2} \bigtriangledown 
({\bf \Omega}\times {\bf r})^2
     = { \Omega}^2{\bf r}_\parallel      
 ;  \hspace {1cm}  
  {\bf h} = \bigtriangledown \times ({\bf \Omega}\times {\bf r})
       = 2{\it c}{\bf \Omega} \eqno (21)   $$

\noindent
where {\bf r}$_\parallel$ = {\it x}{\bf i} + {\it y}{\bf j}.
 $Eq.$(21) gives the centrifugal force, 
${\bf F}_a$ = {\it m}${\bf g}$ = {\it m}${\Omega}^2{\bf r}_\parallel$,
and the Coriolis force, ${\bf F}_b$ = {\it m}{\bf u} $\times$ {\bf h}/c 
= 2{\it m}{\bf u}$\times {\bf \Omega}$ in the rotating system. 
Therefore, the Coriolis force is produced by
a magnetic-type gravitational field. M\o ller's$^{[5]}$ 
derived result is only approximate because he did not explicitly 
choose the coordinate conditions.

\vskip 0.5cm

{\bf 5. Sagnac Effect}

In a sagnac interferometer, the light beam coming from a source is split 
into two sub-beams.  One sub-beam circulates a loop in a clockwise 
direction, and another one circulates the same loop in a 
counter-clockwise direction. When the whole interferometer with light
source and fringe detector is set in rotation with a rate of 
$\Omega$ rad/sec, a fringe shift $\Delta$ with respect to the fringe
position for the stationary interferometer is observed, as first reported 
by Sagnac in 1913$^{[8,9]}$. The formula is given as

   $$   \Delta = 4 {\Omega} Af/c^2   \eqno (22)  $$

We now derive the above formula simply using the 4-D
element in $Eq.$ (20).  As is well known, the propagation of light has 
zero value of the 4-D line element, that is {\it ds} = 0. If a light beam 
travels along a circle with radius {\it R} in a uniformly rotating
system, $Eq.$ (20) can be simplified as follows: {\it d}{\bf r}
={\it Rd}{$\theta$}{${\bf \theta}_o$}, ${\bf{\Omega}} \times \bf r$ =
$\Omega R$ ${\bf \theta}_o$, then we have

 $$  {ds}^2 = (c^2 - ({\Omega}R)^2)dt^2
  - 2({\Omega}R^2)d{\theta}dt - R^2(d\theta)^2 = 0     \eqno (23)  $$

Solving the quadratic equation (23) for $Rd{\theta}/dt$, we obtain the 
speed of light in the frame of the rotating disk:

   $$   c_{\theta} = Rd{\theta}/dt = c \mp {\Omega}R  
 \eqno (24)  $$

\noindent
The sign of minus or plus in $Eq.$(24) depends on the direction
of the light beam. When one beam circulates a loop in a clockwise 
direction, and another one circulates in the opposite 
direction, the time difference is
   
   $$    {\Delta}t = \frac{2{\pi}R}{c-{\Omega}R} -  
   \frac{2{\pi}R}{c+{\Omega}R} \simeq 4{\pi}R^2{\Omega}/c^2 
 \eqno (25)  $$

Since the shift of fringes equals ${\Delta}t$ multiplying
the frequency of the light, $Eq.$(25) leads to  ${\Delta}$=
4${\Omega}Af/c^2$, which is the same as $Eq.$(22).  

As a matter of fact, the Sagnac effect has clearly shown that
(i). one-way speed of light is direction-dependent
 in a rotating system, which is given in $Eq.$(24);
(ii). the transformation of space and time coordinates between
an inertial system and a rotating system obeys the infinitesimal 
Galilean transformations, instead of the Lorentz transformations. 

\vskip 0.5cm
\noindent
{\bf 6. Brief Discussion on the gravitodynamic Field Equations} 

In this paper, the term ``gravitodynamics" 
is used as a replacement for the term ``gravitation" since 
there are similarities between gravitational theory and 
electrodynamics. For instance, in the systems 
with weak gravitational fields and low mass velocities ({\it u}
$\ll $ {\it c}), Einstein's field equations can be  written 
in a form analogous to Maxwell's equations:$^{[10-14]}$

 $$ \bigtriangledown \cdot {\bf g} = - 4 \pi G \rho _m \eqno (26a) $$

 $$ \bigtriangledown \times {\bf h} = 
N(- \frac{4\pi G}{c} \rho _m  {\bf u} + 
\frac{1}{c} \frac{\partial {\bf g}}{ \partial t})   \eqno (26b)  $$

 $$ \bigtriangledown \times {\bf g} = - \frac {1}{c} 
\frac {\partial {\bf h}}{\partial t}    \eqno (26c)  $$
 $$ \bigtriangledown \cdot {\bf h} = 0  \eqno (26d)  $$

\noindent
 In Eqs.(26), 4-D space-time is sliced into 3-D space plus 1-D time
(3+1 split).  
Because of the different approach employed to approximate
Einstein's field equations, the deduced Maxwell-type
field equations in Ref.[10-14] have small differences
with Eqs.(26) in high order terms. For instance, the right
hand side term in (26c) is  a high order term, which is
often set to zero.  
Regardless of these small differences,
 Thorne$^{[10]}$ pointed out that there are three major differences
between Eqs.(26) and 
Maxwell's equations: (i) minus signs in the source term of Eqs.(26a) 
and (26b) are due to the fact that
the gravitational field is attractive rather than repulsive;
(ii) a factor {\it N}=4
 on the right hand side of $Eq.$(26b) presumably is due to 
the fact that gravitational
potential {\it g}$_{{\mu}{\nu}}$ has ten independent components;
(iii) a charge density is replaced by mass density $\rho_m$
times Newton's gravitational constant. 
Furthermore, Thorne emphasized the practical importance of
such ``3 + 1 split" for experimental physicists. As a matter of
fact, preparations are already underway for experiments to 
search magnetic-type gravitational field.$^{[10,11,15]}$
In principle, the properties of gravitational waves can be derived 
from the field equation (26). However, derivation of the
velocity of gravitational waves should require using  
 rigorous forms of the gravitodynamic field equations. As an
example, it is easily seen that the value of $N$ in $Eq$.(26)
would affect the velocity of gravitational waves.

Following the regular procedures to derive   
Einstein's gravitational field equations, one could apply 
the variation principle:
 
  $$   \delta (S_{mf} + S_{f}) = {\delta} \int ({\cal L}_{mf}
 + {\cal L}_f)dVdx^o = 0   \eqno (27) $$               

\noindent
where $dVdx^o$=$dxdydzdx^o$ is 4-D volume element.
${\cal L}_{mf}$ is the part of Lagrangian density that depends on 
the interaction between particles and gravitodynamic field, while
${\cal L}_f$ is the part that depends on 
the gravitodynamic field itself. Both of them could be derived
from $Eq.$(6) and its metric tensor. 

We emphasize that the metric tensor in our formulation
has only four non-constant components, 
${\Psi _o}$ and $\bf {\Psi}$. Therefore the derived
gravitodynamic field equations would have four component equations
(3+1 split), instead of ten equations as in general relativity. 
For this reason, the gravitodynamic field equations 
for a weak field would reduce to
a form analogous to Maxwell's equations. Such equations 
are approximately shown in Eqs.(26) with a constant 
{\it N}=1 in our formulation.

Generally speaking, the rigorous ``3 + 1 split" form of 
the gravitodynamic field equations should be a set of non-linear 
differential equations in terms of  ${\Psi _o}$,
 $\bf {\Psi}$, {\bf g} and {\bf h}.
This approach deserves further investigation.
Furthermore, since the gravitational equation of motion and 
gravitodynamic field equations in our formulation
can be written in 3-D vector forms, one still has freedom
to choose 3-D curvilinear spatial coordinates in a 
given reference system. At this stage, the process of general
covariance has its limited application. 

\vskip 0.4cm
\begin{center}
{\bf Acknowledgements}
\end{center}

\medskip
The author thanks Dr. J.Huang and Dr. M.Hickey for valuable discussions.

\newpage
\vskip 0.4cm
\noindent
{\bf Appendix}

\vskip 0.4cm
\noindent
{\bf A Derivation of the Gravitodynamical Field Equations} 

In the following, a possible form of the gravitodynamical field equations
will be derived from $Eq.$(27). First, ${\cal L}_{mf}$ can be 
obtained from $Eq.$(11).

 $$  {\cal L}_{mf} = (\rho_m{c}^2){\sqrt{{\Psi_o} -  
2{\bf \Psi}\cdot {\bf u}/c  - u{^2}/c{^2}}} \eqno  (28) $$

\noindent
For the weak field and low mass velocities, $Eq$.(28) can be
approximately rewritten as

 $$  {\cal L}_{mf} = -\rho_m(-c^2 + u{^2}/2 - {c^2}{\phi} +  
c({\bf \Psi}\cdot {\bf u}))   \eqno  (29) $$

\noindent
$Eq$.(29) is quite similar to the  ${\cal L}_{mf}$(EM) in
electrodynamics.

Secondly, regarding the function of ${\cal L}_{f}$, 
we need to consider the
differences between the ``actual" gravitational field and 
the ``equivalent" gravitational field produced by acceleration
in a non-inertial system. Here we only intend to derive the field 
equations for the ``actual" gravitational field produced by
gravitational mass.  For this purpose, we should study
the transformations of field intensities here. 

Start with $Eq$.(9), the transformations of potentials 
${\Psi_o}$ and ${\bf \Psi}$.  It can be rewritten as

 $$ {\phi}' = {\phi} - \frac{1}{2}{\frac{v^2}{c^2}} - 
   \frac{\bf v}{c} \cdot {\bf \Psi}  ;
\hspace {1cm} {\bf \Psi}' = {\bf \Psi} + \frac{\bf v}{c}  
   \eqno (30) $$

\noindent 
where ${\bf v}$ is the relative velocity of 
system $K'$ with respect to system $K$.  Since we only consider the 
actual gravitational field in both systems, a constant  ${\bf v}$ 
is assumed here.  Using the definition of $\bf g$ and $\bf h$ in 
$Eq$.(14), $Eq$.(30) leads to

  $$    {\bf g'} = {\bf g} + {\bf \beta} \times {\bf h};
\hspace {1cm}    {\bf h'} = {\bf h}        \eqno (31) $$

\noindent 
Therefore, the magnetic-type gravitational intensity is an invariant
 3-D vector.  From $Eqs$.(30) and (31), it is easy to prove another
invariant 3-D vector {\bf g}, which is defined as 

    $$ \tilde{\bf g} = {\bf g} - {\bf \Psi} \times {\bf h} = 
                 {\bf g}' - {\bf \Psi}' \times {\bf h}' 
        \eqno (32) $$

In terms of the above invariant, various kinds of ${\cal L}_f$
can be introduced.  Comparing the function of ${\cal L}_f$(EM)
in electrodynamics, we choose the following ${\cal L}_f$ for
discussion:

 $$  {\cal L}_{f} = -(8\pi G)^{-1}({\tilde{\bf g}} \cdot {\tilde{\bf g}}
     - {\bf h \cdot h})   \eqno  (33) $$

\noindent 
Substituting $Eq$.(32) into (33), and neglecting the high order terms, we 
obtain

  $$  {\cal L}_{f} = -(8\pi G)^{-1}({\bf g \cdot g} - 2{\bf g}
    \cdot ({\bf \Psi} \times {\bf h}) -   
   {\bf h \cdot h})   \eqno  (34)   $$

Substituting the expressions of ${\cal L}_{mf}$
in $Eq$.(29) and ${\cal L}_{mf}$ in $Eq$.(34) into 
the $Eq$.(27), and following the same procedures 
to derive the field equations 
in electro-dynamics [7], we can obtain two 
gravitodynamic field equations:

 $$ \bigtriangledown \cdot ({\bf g - \Psi \times h}) 
= - 4 \pi G \rho _m              \eqno (35a) $$

 $$ \bigtriangledown \times ({\bf h - \Psi \times g}) = 
- \frac{4\pi G}{c} {\rho _m}{\bf u} + 
\frac{1}{c} \frac{\partial}{\partial t}
({\bf g - \Psi \times h})   \eqno (35b)  $$

\noindent
When comparing $Eq.(35a,b)$ with Maxwell's equations,
it is easy to see that there are additional terms,
${\bf \Psi \times h}$ and ${\bf \Psi \times g}$, 
in $Eq.(35a,b)$.

Using the mathematical properties of the operator 
$\bigtriangledown$, other two gravitational field equations
can be easily derived from $Eq$.(14):

 $$ \bigtriangledown \times {\bf g} = - \frac {1}{c} 
\frac {\partial {\bf h}}{\partial t}    \eqno (35c)  $$
 
  $$ \bigtriangledown \cdot {\bf h} = 0  \eqno (35d)  $$

\noindent
Furthermore, from $Eq$.(35a) and $Eq$.(35b), the equation of 
continuity can be obtained 

  $$   \frac{1}{c} \frac{\partial {\rho _m}}{\partial t}
  + \bigtriangledown \cdot ({\rho _m}{\bf u}) = 0  \eqno (36)  $$


\newpage

\begin{thebibliography}{99}

\bibitem{ }J. D. Norton, ``General Covariance and the Foundations
of General Relativity", Rep. Prog. Phys., {\bf 56}, no.7, 791 (1993).

\bibitem{ }A. Einstein, ``The Foundation of the General
Theory of Relativity", Ann. Physik., {\bf 49}, 769 (1916).

\bibitem{ } A. Einstein, ``The Meaning of Relativity",
Princeton Univ. Press, New Jersey, Chap.3 (1955).

\bibitem{ }Y. Duan {\it et al}, ``A New Coordinate Condition 
in General Relativity", Gen. Rel. and Gravi., {\bf 24},
1033 (1992).

\bibitem{ }C. M\o ller, ``The Theory of Relativity", 
 Second edition, Clarendon Press, Oxford, Chap.8-11 (1972).

\bibitem{ }S. Weinberg, ``Gravitation and Cosmology", 
John Wiley and Sons Inc., New York, Chap.3 (1972).

\bibitem{ }L. D. Landau and E. M. Lifshitz, ``The 
Classical Theory of Fields",
Fourth English Edition, Pergamon Press, Chap.3-4, 10-12 (1975).

\bibitem{ }G. Sagnac, Compt. Rend., {\bf 157}, 
    (1913)708, 1410.

\bibitem{ }J. E. Post, ``Sagnac Effect", Rev. Mod. Physics, {\bf 39}, 
    (1967)475.

\bibitem{ }K. S. Thorne, 
 ``Gravitomagnetism, Jets in Quasars, and the
Stanford Gyroscope Experiment", in the book `` Near Zero:
New Frontiers of Physics" by J. D. Fairbank {\it et al},
New York, P.573 (1988).

\bibitem{ }N. Li and D. G. Torr, ``Effect of a gravitomagnetic 
field on pure superconductors",  Phys. Rev. D., {\bf 43}, 457 (1991).

\bibitem{ }V. B. Braginsky, C. M. Caves, K. S. Thorne, 
``Laboratory Experiments to Test Relativistic Gravity"",
 Phys. Rev. D., {\bf 15}, 2047 (1977).

\bibitem{ }E. G. Harris, ``Analogy between general relativity and 
electromagnetism",  Am. J. Phys., {\bf 59}, 421 (1991).

\bibitem{ }T. Chang, ``Imaginary Charge and Gravitational-
Electric Space", Galilean Electrodynamics,
{\bf 3}, No.2, 36 (1992).

\bibitem{ } C. Will, ``Theory and Experiment in
Gravitational Physics",
Cambridge Univ. Press, New York, Chap.9 (1981).

\end{thebibliography}
\end{document}